\documentclass[useAMS,usenatbib,usegraphicx]{mn2e}

\newcommand{\msun}{M$_\odot$}
\newcommand{\lsun}{L$_\odot$}

\newcommand{\mc}{\multicolumn}

\title[SN rate in local galaxy clusters] 
{The supernova rate in local galaxy clusters}

\author[Mannucci et al]{
F. Mannucci$^1$\thanks{E-mail:filippo@arcetri.astro.it},
D. Maoz$^{2,3}$,
K. Sharon$^2$,
M. T. Botticella$^4$,
M. Della Valle$^{5,3,6}$,
\newauthor
A. Gal-Yam$^{7,8}$, and
N. Panagia$^{9,10,11,6}$\\
$^1$ INAF - Istituto di Radioastronomia, Largo E. Fermi 5, 
      50125 Firenze, Italia\\
$^2$ School of Physics and Astronomy, 
      Tel-Aviv University, Tel-Aviv 69978, Israel\\
$^3$ Kavli Institute for Theoretical Physics, 
     University of California, Santa Barbara, CA 93106-4030, USA\\
$^4$ INAF - Osservatorio Astronomico di Collurania, Teramo, Italia\\
$^5$ INAF - Osservatorio Astrofisico di Arcetri, Firenze, Italia\\
$^6$ ICRANET, Piazzale della Repubblica 10, I-65122, Pescara, Italy\\
$^7$ Dept. of Astronomy, California Institute of Technology, 
      Pasadena, CA 91125, USA\\
$^8$ the Astrophysics group, Weizmann Institute of Science, 
        Rehovot, 76100, Israel\\
$^9$ STScI, 3700 San Martin Drive, Baltimore, MD 21218, USA\\
$^{10}$ INAF, - Osservatorio Astrofisico di Catania,
Via S. Sofia 78, I-95123 Catania, Italy\\
$^{11}$ Supernova Ltd., Olde Yard Village \#131, Northsound Road, Virgin Gorda,
}

\begin{document}
\date{Submitted; Accepted}
\pagerange{\pageref{firstpage}--\pageref{lastpage}} \pubyear{2007}
\maketitle

\begin{abstract}

We report a measurement of the supernova (SN) rates (Ia and core-collapse)
in galaxy clusters based on the
136 SNe of the sample described in \citet{C99} and \citet{M05}.

Early-type cluster galaxies
show a type Ia SN rate (0.066 SNuM) similar to that
obtained by \citet{sharon07} and more than 3 times larger
than that in field early-type galaxies (0.019 SNuM). 
This difference has a 98\% statistical confidence level.
We examine many possible observational biases which could affect the rate
determination,
and conclude that none of them is likely to significantly alter the results.
We investigate how the rate is related to 
several properties of the parent galaxies, and find that cluster membership,
morphology and radio power all affect the SN rate, while galaxy mass
has no measurable effect. 
The increased rate may be due to galaxy interactions in clusters, inducing
either the formation of young stars or a different evolution of 
the progenitor binary systems.

We present the first measurement of the core-collapse SN rate in cluster
late-type galaxies, 
which turns out to be comparable to the rate in field galaxies.
This suggests that no large systematic difference in the initial mass
function exists between the two environments. 
\end{abstract}


\section{Introduction}
\label{sec:intro}

Type Ia Supernovae (SNe Ia) are believed to be the result of the thermonuclear
explosion of a C/O white dwarf (WD) in a binary system due to mass exchange
with the secondary star. 
This conclusion follows from a few fundamental
arguments: the explosion requires a degenerate system, such as a white dwarf;
the presence of SNe Ia in old stellar systems implies that at least some of
their progenitors must come from old, low-mass stars; the lack of hydrogen
in the SN spectra requires that the progenitor has lost its outer envelope;
and, the released energy
per unit mass is of the order of the energy output of the 
thermonuclear conversion of carbon or oxygen into iron. 
Considerable uncertainties about the explosion model
remain within this broad framework, such as the structure and
the composition of the exploding WD (He, C/O, or O/Ne), the mass at
explosion (at, below, or above the Chandrasekhar mass) and the flame 
propagation (detonation, deflagration, or a combination of the two).
The key observations constraining the explosion
models are the light curve and the evolution of the spectra.

Large uncertainties also remain regarding
the nature of the progenitor binary system,
its evolution through one or more common envelope phases, and
its configuration (single or double-degenerate)
at the moment of the explosion (see \citealt{yungelson05}, for a review).
Solving the problem of the progenitor system is of great importance for modern
cosmology as SNe dominate metal production,
(e.g., \citealt{matteucci86}), are expected to be important producer
of high-redshift dust \citep{maiolino01, maiolino04a,maiolino04b,bianchi07},
and are essential to understand the feedback process during galaxy 
formation (e.g., \citealt{scannapieco06}).
The nature of the progenitor systems can be probed by studying the 
SN rate in different stellar populations, and constraining the delay time
distribution (DTD) between star formation and SN explosion.

\smallskip

In 1983, Greggio \& Renzini computed the expected DTD for a single-degenerate
system. The computation was later refined by many authors
and extended to double-degenerate systems
\citep{tornambe86,tornambe89,tutukov94,yungelson00,matteucci01,belczynski05,greggio05}.
The DTD can be convolved with
the star formation history (SFH) of each galaxy to obtain its SN rate.
The observation of the SN rates per unit mass in galaxies of different types
\citep{M05,sullivan06}
and in radio-loud early-type galaxies \citep{dellavalle05}
has proved to be an effective way to constrain the DTD.
The SN rates per unit mass show that SNe Ia must come from both young 
and old progenitors \citep{M05,sullivan06}.
The dependence of the SN rate on the radio power of the parent galaxy
is well reproduced by a
``two channel'' model \citep{mannucci06},
in which about half of the SNe Ia, the so-called
``prompt'' population, explode soon
after the formation of the progenitors, on time scales shorter than 
$10^8$ yr,
while the other half (the ``tardy'' population)
explode on a much longer time scale, of the order of $10^{10}$ yr.
Several attempts to compare the evolution of SN rate with redshift with
that of the SFR have also been presented 
(see, among many others, 
\citealt{galyam04,dahlen04,cappellaro05,neill06,barris06,botticella07} and
\citealt{poznanski07}),
but the large uncertainties on both quantities prevent strong conclusions
(see, for example, \citealt{forster06}). 

\smallskip

In principle, an accurate measurement of the
DTD could identify the progenitor binary system. In practice, both
the large number of free parameters involved in the theoretical computations
of the DTD,
and the complex SFHs of most of the galaxies 
make this identification much more uncertain.
To solve the problem of the complexity of the SFH, it is
interesting to measure the SN Ia rate in galaxy clusters.
Most of the stellar mass of these systems is contained in elliptical 
galaxies, whose stellar populations are dominated by old stars. 
Despite the problem that even a small amount of new stars could give
a significant contribution to the SN rate (see the discussion in 
sect.~\ref{sec:discussion}), the reduction in the uncertainty in the SFH
is of great help to derive the DTD. 

\smallskip

The cluster SN rate is also of great importance to study the
metallicity evolution of the universe.
The gravitational potential well of galaxy clusters is deep enough to retain
in the intracluster medium (ICM) all the metals which are produced in 
galactic or intergalactic SNe.
As a result, the metallicity of the ICM
is a good measure of the integrated past history
of cluster star formation and metal production. 
As discussed by \citet{renzini93}, the measured 
amount of iron is an order of magnitude too high to be produced 
by SNe Ia exploding at the current rate. 
Explanations of this effect include the presence of higher SN
rates in the past \citep{matteucci06}, the importance of the intracluster
stellar population \citep{zaritsky04}, or evolving properties of
star formation processes \citep{maoz04,lowenstein06}.
The observed abundance ratios in the ICM can be used
to constrain the ratio between the total numbers of Ia and CC SNe,
as recently done by \citet{deplaa07}.
Constraints on the SN Ia models can also be derived from the
radial distribution of metallicity 
\citep{dupke02}.
\citet{calura07} used the observed cosmic evolution
of iron abundances in \citet{balestra07} to constrain the history
of SN explosion, iron formation and gas stripping in galaxy clusters. 
They found good
agreement with the observations, especially when the ``two channel'' model
of SNe Ia by \citet{mannucci06} is used.

\smallskip

There are strong motivations for measuring also the cluster rates of the other
physical class of SNe, the core-collapse (CC) group.
Type II and type Ib/c SNe are attributed to this group because
there is a general consensus that these explosions are due to
the collapse of the core of a massive (about 8--40~\msun) star.
Thus, CC SNe are expected to be good tracers of star formation
in moderately dusty environments (see \citealt{mannucci07}).
Their rate per unit mass is also very sensitive to the initial mass function 
(IMF), because
SN explosions are due to massive stars while most of the mass
is locked in low-mass stars. As a consequence, studying the CC SN rate as a
function of environment is a sensitive test for any systematic
difference in IMF.

\begin{table}
\caption{ Measured type Ia SN rates in early-type cluster galaxies
\label{tab:history}
}
\begin{tabular}{llcc}
\hline
\hline
Reference         & ~~$z$ ~~($z$ range)  & N$_{SN}$& Rate\\
                  &                     &         & (SNuB) \\
\hline
This work         & 0.02 (0.005--0.04)  &  20 & $0.28^{+0.11}_{-0.08}$\\
\citet{crane77}   & 0.023 (0.020--0.026)&   8 & $\sim$0.10  \\  
\citet{barbon78}  & 0.023 (0.020--0.026)&   5 & $\sim$0.16  \\  
\citet{germany04} & 0.05 (0.02--0.08)   &  23 & unpubl. \\
\citet{sharon07}  & 0.15 (0.06--0.19)   &   6 & 0.27$^{+0.16}_{-0.11}$\\
\citet{galyam02}  & 0.25 (0.18--0.37)   &   1 & 0.39$^{+1.65}_{-0.37}$\\
\citet{galyam02}  & 0.90 (0.83--1.27)   &   1 & 0.80$^{+0.92}_{-0.41}$\\
\hline
\end{tabular}
\end{table}

\subsection{The observed cluster supernova rate}

Prompted by all these motivations, several groups have measured the
SN Ia rate in galaxy clusters, but the results are still quite sparse. 
The first published values are due to
\citet{crane77} and \citet{barbon78}
(see Table~\ref{tab:history} for a summary, including the results of 
our work, discussed below), 
before 
a clear distinction between type Ia and Ib/c had been introduced. 
They used a sample of 5--8 SNe in the Coma Cluster
and constrained the SN rate to be of the order of 0.15 SNuB 
(SN per century per $10^{10}$ \lsun\ in the B band). 
The SN rate as a function of galaxy environment
was also addressed by 
\citet{caldwell81} to derive information on SN progenitors.

Modern searches for cluster SNe begin with \citet{norgaard89}
who discovered a SN Ia in a cluster at $z=0.31$.
Starting from the late '90s, the Mount Stromlo 1.3 m telescope was used to
monitor a few tens of Abell Clusters \citep{reiss98}. Three years of
monitoring resulted in the detection of 23 candidate SNe Ia in cluster galaxies
\citep{germany04}, but a rate based on this sample was never published.

The first rates for cluster galaxies 
based on modern searches were published by 
\citet{galyam02}. These authors used archive images from the
Hubble Space Telescope (HST) 
of 9 galaxy clusters,
and discovered 6 SNe, 2 of which are associated 
with the clusters, at $z=0.18$ and  $z=0.83$. 
The derived rates were affected by large statistical 
uncertainties due to the small number of detected SNe, but were consistent 
with a moderate increase of the rate with redshift
compared to the rate in local elliptical galaxies.
A sample of 140 low-redshift Abell clusters were monitored 
by the Wise Observatory Optical Transient Search (WOOTS, \citealt{galyam07})
using the Wise 1m telescope.
The seven detected cluster SNe were used to
constrain the fraction of intergalactic stars and SNe \citep{galyam03}
and to measure the cluster SN rate \citep{sharon07}.
This latter work obtains a value of the SN rate per unit mass
of $0.098^{+0.058}_{-0.039}$ SNuM (SN per century per $10^{10}$ \msun\ of
stellar mass), which is larger than, but still
consistent with, the value of $0.038^{+0.014}_{-0.012}$ SNuM, derived 
by \citet{M05} for local ellipticals.
Finally, a SN search in clusters is ongoing at the Bok Telescope on Kitt
Peak \citep{sand07}.

All of the previous published SN Ia rates 
are based on a small number of SNe and,
as a consequence, have large statistical errors. 
Also, a cluster rate for CC SNe has never been published
because many of the cited samples only contain Ia SNe.
In this work, 
we use the SNe in the 
\citet{C99} sample to study the SN rate as a
function of galaxy environment.  

Throughout this paper we use the ``737'' values of the cosmological
parameters: $(h_{100},\Omega_m,\Omega_\Lambda)=(0.7,0.3,0.7)$.

\begin{figure}
\includegraphics[width=9cm]{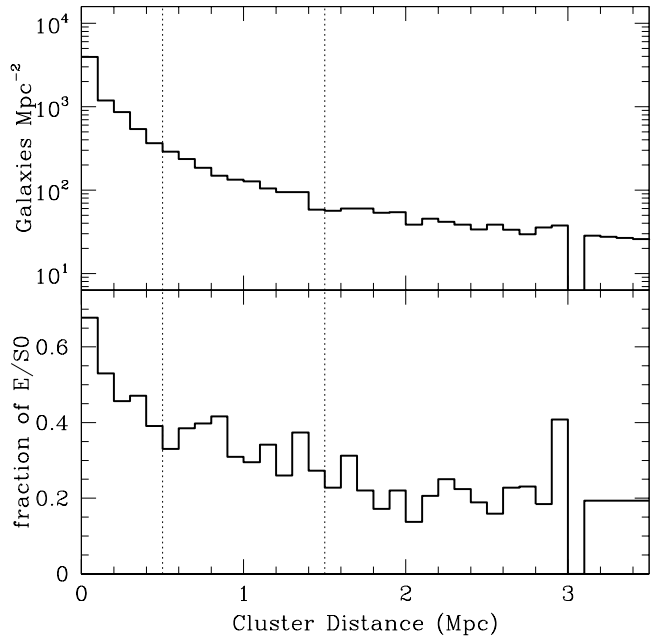}
\caption{
\label{fig:galdist}
Surface density of galaxies (upper panel) and fraction of 
early-type galaxies (lower
panel) as a function of the projected distance from the closest cluster.
Above 3 Mpc, the average for field galaxies is shown.
The vertical dotted lines show the two projected distances, 0.5 and 1.5 Mpc,
used to define cluster galaxies (see text).
}
\end{figure}

\section{Measuring the rates in clusters and in the field}
\label{sec:rate}

The SN sample described by \citet{C99} consists of 136 local SNe 
(with redshifts $z<0.04$),
obtained by monitoring 8349 galaxies for many years.  It is based
on 5 visual and photographic searches and, to date, it comprises the 
largest published sample of SNe suitable for rate measurements. 

The monitored galaxy sample is very heterogeneous and for most of 
the galaxies a
clear membership in a cluster is not known. 
To test for cluster membership of each galaxy 
we used the list of known galaxy clusters in 
the NASA/IPAC Extragalactic Database (NED). 
We considered a galaxy to be part of a cluster if 
its radial velocity is within 1000 km~s$^{-1}$ of that of the 
known cluster, and if
its projected distance is below a certain distance
\citep{dressler97,hansen05}.
Figure~\ref{fig:galdist} shows how such a threshold distance was chosen.
The upper panel shows the surface density of galaxies as a function of
the projected distance $D$ from the closest cluster. 
At $D<1.5$ Mpc the density 
of galaxies shows an increase over the
large distance value ($D>2$ Mpc), and a steepening
of the increase for $D<0.5$ Mpc.
In the lower panel of Fig.~\ref{fig:galdist}, the fraction of early-type
galaxies is shown as a function of $D$.
As expected from the density-morphology 
relation (e.g., \citealt{dressler97,goto04,park07}),  a sharp increase
of this fraction is seen at small distances. 
At large distances , about 19\%
of the galaxies in our sample are early-type 
(defined to have a morphological index T$<$-1.5 in 
the HyperLeda catalog, \citealt{paturel03}) , while this fraction rises to
an average of 43\% at $D<1.5$ Mpc and 53\% at $D<0.5$ Mpc. 
Based on these considerations, we
see that the population within 0.5 Mpc of a cluster are dominated by cluster
members, galaxies more distant then 1.5 Mpc are mostly field galaxies,
and a mixture of the two populations are probably present between these
two distances.
We therefore consider as cluster members the galaxies having $D<0.5$ Mpc,
but throughout the paper we will also discuss the effect of including
all the galaxies with $D<1.5$ Mpc.
Galaxies with $D>1.5$ Mpc will be considered as belonging to the field
population.

\smallskip

The above classification
has a number of weaknesses. First, we assume that all clusters 
have the same radial extent, even if this is known not to be true.
Second, clusters show a galaxy density that smoothly decreases with radius 
rather than a sharp cutoff.
Third, the NED cluster
catalog is not complete and it is possible that some clusters are missing.
All of these effects are likely to produce some degree of misclassification
in both directions. However,
Fig.~\ref{fig:galdist} shows that these effects, if present,
are not strong enough to completely spoil the classification.
Also, any missclassification 
can dilute or hide an existing difference in rates,
but it is unlikely to produce an artificial difference in rates
or enlarge a small difference.

\smallskip

Of the 8349 galaxies in the full sample, 810 (about 10\%)
belong to clusters with $D<0.5$ Mpc (1666 for $D<1.5$ Mpc),
and 6683 galaxies belong to the field.
The expected strong morphological segregation is well recovered:
(1) more early-type galaxies are present in clusters, as noted above;
and (2) late spirals and irregulars (T$>$3.5)
constitute 15\% of the cluster galaxies and 45\% of the field galaxies.
It is not straightforward to compare these fractions to 
those generally observed in complete galaxy samples,
because the relative fractions
depend strongly on cluster radial distance, galaxy overdensity and 
morphological classification. Furthermore, the original galaxy
sample was not selected in order to reproduce the cosmic average
but rather to have a significant number of SN detections. 
Several authors \citep{dressler97,smith05,sorrentino06,park07}
report the morphological mix of galaxies in clusters and in the field,
obtaining that within 1 Mpc, clusters have $\sim$50-70\% early-type galaxies 
(E+S0) and 30-50\% spirals, with opposite fractions in the field.
Our galaxy sample thus seems to be roughly consistent with
the cosmic average. 

Cluster galaxies host 20 of the SNe of our sample (14\% of the total), 
and field galaxies host the remaining 92 SNe. 
Galaxies with $D<1.5$ Mpc host 44 SNe.
As explained in \citet{C99}, 10 out of 136 SNe have incomplete 
classification and have been redistributed among the three basic type of SNe
(Ia, II and Ib/c) according to the observed relative fractions.
As a consequence,
some bins in the distributions discussed below 
contain a fractional number of SNe.

Rates are computed as a function of 
of the B-band luminosity, as in \citet{C99},
and as a function of galaxy stellar mass, as described in \citet{M05},
with mass computed 
from the K-band luminosity and the (B--K) color. 
For each galaxy, the control time (CT, i.e., the time during which
a SN could be detected by the survey) for each
SN type was computed. The ``sensitivity'' to each SN type is proportional to
the product of the CT times either the mass or the B luminosity for the
rates per unit mass and per unit luminosity, respectively.

\smallskip

\begin{figure*}
\includegraphics[angle=-90,width=15cm]{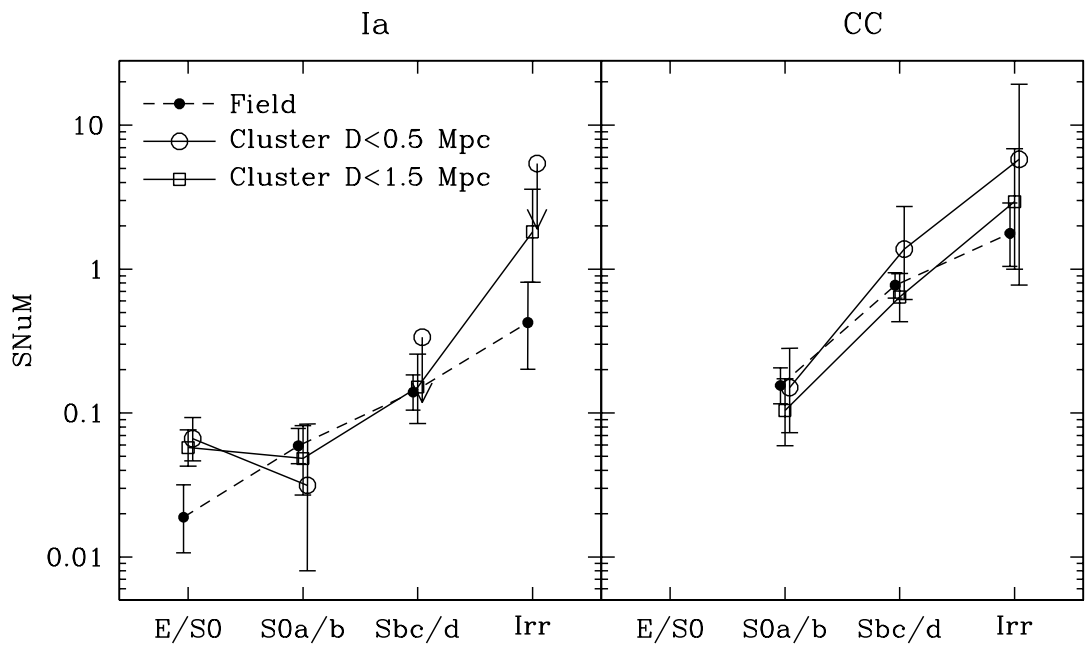}
\caption{
\label{fig:massrate}
SN rates per unit mass as a function of galaxy morphology 
for type Ia ({\em left}) and CC SNe ({\em right}). 
Error bars and upper limits on the rate
correspond to a 1$\sigma$ confidence level.
}
\end{figure*}

\begin{figure*}
\includegraphics[angle=-90,width=15cm]{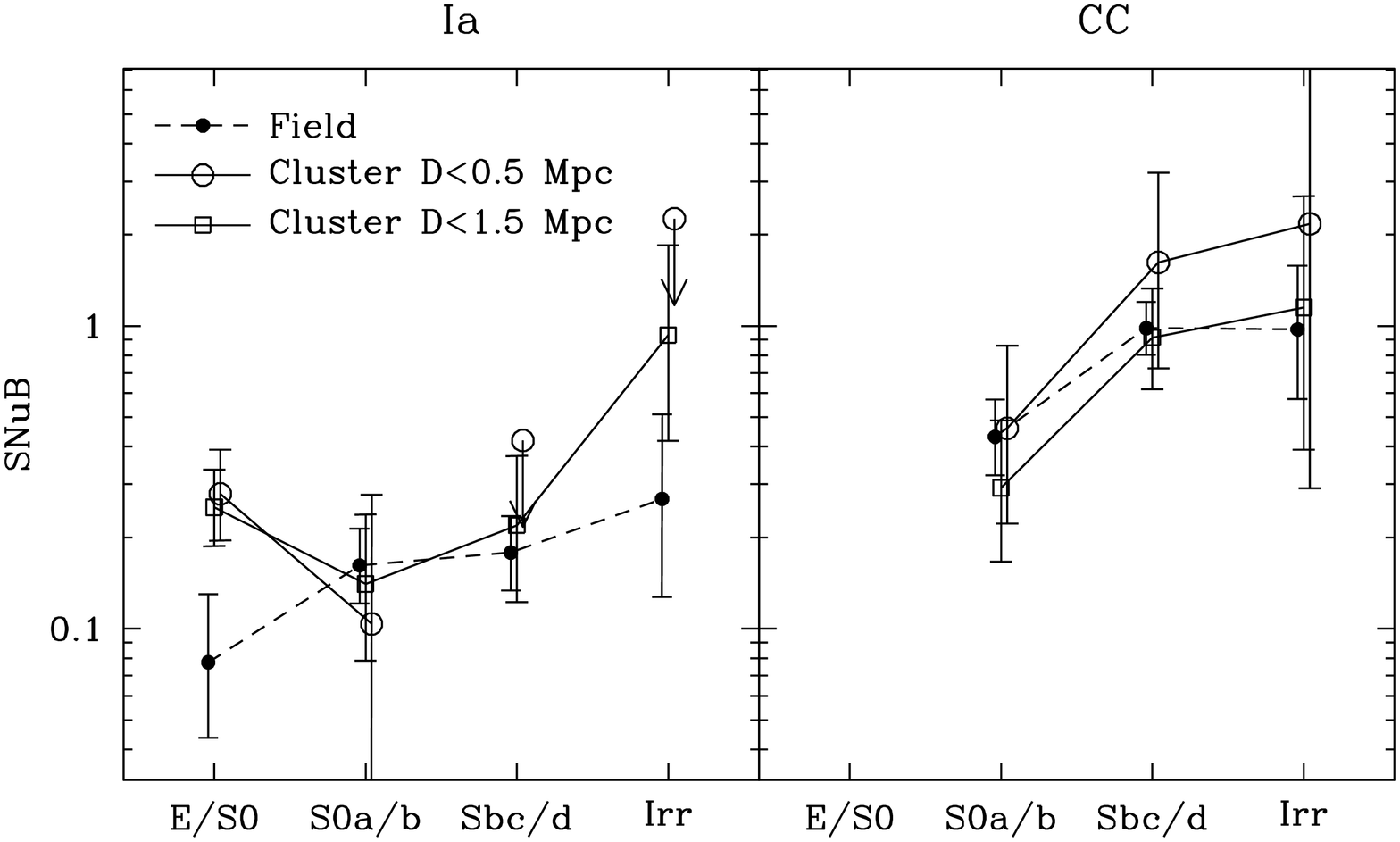}
\caption{
\label{fig:brate}
SN rates per unit B-band luminosity as a function of galaxy 
morphology for type Ia ({\em left}) and CC SNe ({\em right}).
}
\end{figure*}

Table~\ref{tab:massrate} and Fig.~\ref{fig:massrate} 
show the SN rate per unit mass in clusters (both for $D<0.5$~Mpc and 
for $D<1.5$~Mpc)
and in the field
as a function of the morphology of the parent galaxy, 
with 1$\sigma$ statistical errors.
Table~\ref{tab:brate} and Fig.~\ref{fig:brate}
show the corresponding rates per unit B luminosity. 
Tables~\ref{tab:massrate} and \ref{tab:brate} also list the
total rate without binning by morphology. For CC SNe, 
the two classes Ib/c and II are given, together with their sum (labeled CC). 

In the next two sections, we discuss these results, first for the CC SNe
(Sect.~\ref{sec:ccrate}), and then for SNe Ia (Sect.~\ref{sec:iarate}).

\begin{table*}
\caption{
Cluster and field SN rate per unit mass as a function of morphology. 
For each SN type the number of SNe and the rate in SNuM are given. 
Ngal is the number of galaxies per morphological bin.
\label{tab:massrate}
}
\begin{tabular}{crrrrrrrrr}
\hline
\hline
Type  & Ngal&     \mc{2}{c}{Ia}    &  \mc{2}{c}{Ib/c}  &   \mc{2}{c}{II}    &  \mc{2}{c}{CC}      \\
      &     &  N  &      Rate~~~~~ &  N  &    Rate~~~~~  &  N  &      Rate~~~~~ &  N  &    Rate~~~~~  \\
\hline
      \mc{10}{c}{Cluster $D<0.5$ Mpc} \\
\hline
E/S0 & 430& 11.0& $0.066^{+0.027}_{-0.020}$&  0.0 & $<0.020$                 & 0.0 & $<0.027$                  &  0.0 & $<0.047$                 \\ 
S0a/b& 251&  1.5& $0.031^{+0.052}_{-0.023}$&  2.5 & $0.094^{+0.105}_{-0.056}$& 1.0 & $0.056^{+0.131}_{-0.049}$ &  3.5 & $0.150^{+0.132}_{-0.077}$ \\ 
Sbc/d& 100&  0.0& $<0.34$                  &  0.0 & $<0.64$                  & 3.0 & $1.37 ^{+1.34 }_{-0.76 }$ &  3.0 & $1.376^{+1.347}_{-0.761}$ \\ 
Irr  &  29&  0.0& $<5.4$                   &  0.0 & $<7.3$                   & 1.0 & $5.79 ^{+13.4  }_{-5.01 }$&  1.0 & $5.79 ^{+13.4  }_{-5.01 }$ \\ 
\hline
TOT &  810& 12.5& $0.057^{+0.021}_{-0.016}$&  2.5 & $0.020^{+0.022}_{-0.012}$& 5.0 & $0.057^{+0.039}_{-0.025}$ &  7.5 & $0.077^{+0.040}_{-0.028}$ \\
\hline
      \mc{10}{c}{Cluster $D<1.5$ Mpc} \\
\hline
E/S0 & 723& 15.0& $0.058^{+0.019}_{-0.015}$& 0.0 & $<0.012$                  &  0.0 & $<0.017$                  &  0.0 & $<0.029$                  \\ 
S0a/b& 519&  4.8& $0.048^{+0.034}_{-0.021}$& 2.7 & $0.044^{+0.047}_{-0.026}$ &  2.5 & $0.061^{+0.068}_{-0.036}$ &  5.2 & $0.104^{+0.069}_{-0.045}$ \\ 
Sbc/d& 321&  4.8& $0.152^{+0.105}_{-0.067}$& 0.6 & $0.031^{+0.105}_{-0.031}$ &  8.5 & $0.610^{+0.290}_{-0.206}$ &  9.1 & $0.641^{+0.290}_{-0.209}$ \\ 
Irr &   94&  3.0& $1.82 ^{+1.78}_{-1.00}$  & 1.0 & $1.28^{+2.96}_{-1.10}$ &  1.0 & $1.67^{+3.88}_{-1.45}$ &  2.0 & $2.94^{+3.91}_{-1.94}$ \\ 
\hline
TOT & 1666& 27.7& $0.070^{+0.016}_{-0.013}$& 4.3 & $0.018^{+0.014}_{-0.008}$ & 12.0 & $0.072^{+0.028}_{-0.021}$ & 16.3 & $0.090^{+0.028}_{-0.022}$ \\
\hline
      \mc{10}{c}{Field} \\
\hline
E/S0 & 1326&  5.0& $0.019^{+0.013}_{-0.008}$& 0.0 & $<0.015$                  &  0.0 & $<0.020$                  &  0.0 & $<0.034$                  \\ 
S0a/b& 2393& 15.7& $0.059^{+0.019}_{-0.015}$& 3.3 & $0.026^{+0.023}_{-0.014}$ & 12.0 & $0.130^{+0.049}_{-0.037}$ & 15.3 & $0.155^{+0.051}_{-0.039}$ \\ 
Sbc/d& 2362& 15.8& $0.140^{+0.045}_{-0.035}$& 5.8 & $0.121^{+0.075}_{-0.049}$ & 23.5 & $0.652^{+0.164}_{-0.134}$ & 29.2 & $0.773^{+0.171}_{-0.142}$ \\ 
Irr  &  551&  3.3& $0.426^{+0.38}_{-0.22}$& 1.1 & $0.300^{+0.62}_{-0.25}$ &  4.5 & $1.47^{+1.08}_{-0.67}$ &  5.7 & $1.77^{+1.11}_{-0.73}$ \\ 
\hline
TOT  & 6683& 39.8& $0.061^{+0.011}_{-0.010}$&10.2 & $0.033^{+0.014}_{-0.010}$ & 40.0 & $0.174^{+0.032}_{-0.027}$ & 50.2 & $0.207^{+0.034}_{-0.029}$\\ 
\hline
\end{tabular}
\end{table*}

\begin{table*}
\caption{Cluster and field SN rate per unit B-band luminosity
as a function of morphology. For each SN
type the number of SN and the rate in SNuB are given. Ngal is the number of
galaxies per morphological bin.
\label{tab:brate}
}
\begin{tabular}{crrrrrrrrr}
\hline
\hline
Type  & Ngal&     \mc{2}{c}{Ia}    &  \mc{2}{c}{Ib/c}  &   \mc{2}{c}{II}    &  \mc{2}{c}{CC}      \\
      &     &  N  &      Rate~~~~~ &  N  &    Rate~~~~~  &  N  &      Rate~~~~~ &  N  &    Rate~~~~~  \\
\hline
      \mc{10}{c}{Cluster $D<0.5$ Mpc} \\
\hline
E/S0 & 430& 11.0& $0.28^{+0.11}_{-0.08}$&  0.0 & $<0.078$              &  0.0 & $<0.11$               &  0.0 & $<0.19$                   \\ 
S0a/b& 251&  1.5& $0.10^{+0.17}_{-0.08}$&  2.5 & $0.29^{+0.32}_{-0.17}$&  1.0 & $0.17^{+0.40}_{-0.15}$&  3.5 & $0.46^{+0.40}_{-0.24}$ \\ 
Sbc/d& 100&  0.0& $<0.42$               &  0.0 & $<0.74$               &  3.0 & $1.62^{+1.59}_{-0.90}$&  3.0 & $1.62^{+1.59}_{-0.90}$ \\ 
Irr &  29 &  0.0& $<2.2 $               &  0.0 & $<2.7 $               &  1.0 & $2.17^{+5.04}_{-1.88}$&  1.0 & $2.17^{+5.04}_{-1.88}$ \\ 
\hline
TOT & 810 & 12.5& $0.211^{+0.078}_{-0.059}$&  2.5 & $0.070^{+0.078}_{-0.042}$&  5.0 & $0.20^{+0.14}_{-0.09}$&  7.5 & $0.27^{+0.14}_{-0.10}$ \\ 
\hline
      \mc{10}{c}{Cluster $D<1.5$ Mpc} \\
\hline
E/S0 &  723& 15.0& $0.25^{+0.08}_{-0.06}$ & 0.0 & $<0.050$                  & 0.0 & $<0.071$                  &  0.0 & $<0.12$                   \\ 
S0a/b&  519&  4.8& $0.14^{+0.10}_{-0.06}$ & 2.7 & $0.12^{+0.13}_{-0.07}$ & 2.5 & $0.17^{+0.19}_{-0.10}$ &  5.2 & $0.29^{+0.19}_{-0.13}$ \\ 
Sbc/d&  321&  4.8& $0.22^{+0.15}_{-0.10}$ & 0.6 & $0.046^{+0.15}_{-0.04}$ & 8.5 & $0.87^{+0.41}_{-0.29}$ &  9.1 & $0.92^{+0.41}_{-0.30}$ \\ 
Irr  &   94&  3.0& $0.93^{+0.91}_{-0.52}$ & 1.0 & $0.47^{+1.10}_{-0.41}$ & 1.0 & $0.67^{+1.57}_{-0.58}$ &  2.0 & $1.151^{+1.53}_{-0.76}$ \\ 
\hline
TOT  & 1666& 27.7& $0.23^{+0.05}_{-0.04}$ & 4.3 & $0.057^{+0.043}_{-0.027}$ &12.0 & $0.23^{+0.09}_{-0.07}$ & 16.3 & $0.29^{+0.09}_{-0.07}$ \\ 
\hline
      \mc{10}{c}{Field} \\
\hline
E/S0 & 1326& 5.0 &$0.077^{+0.053}_{-0.034}$ & 0.0 & $<0.060$                  & 0.0 & $<0.080$                  &  0.0 & $<0.14$                   \\ 
S0a/b& 2393&15.7 &$0.16^{+0.05}_{-0.04}$ & 3.3 & $0.072^{+0.066}_{-0.038}$ &12.0 & $0.36^{+0.14}_{-0.10}$ & 15.3 & $0.43^{+0.14}_{-0.11}$ \\ 
Sbc/d& 2362&15.8 &$0.18^{+0.06}_{-0.04}$ & 5.8 & $0.15^{+0.09}_{-0.06}$ &23.5 & $0.83^{+0.21}_{-0.17}$ & 29.2 & $0.98^{+0.22}_{-0.18}$ \\ 
Irr  &  551& 3.3 &$0.27^{+0.24}_{-0.14}$ & 1.1 & $0.16^{+0.32}_{-0.13}$ & 4.5 & $0.82^{+0.60}_{-0.37}$ &  5.7 & $0.97^{+0.61}_{-0.40}$ \\ 
\hline
TOT  & 6683&39.8 &$0.151^{+0.028}_{-0.024}$ &10.2 & $0.083^{+0.035}_{-0.026}$ &40.0 & $0.44^{+0.08}_{-0.07}$ & 50.2 & $0.52^{+0.08}_{-0.07}$ \\ 
\hline
\end{tabular}
\end{table*}

\section{Core-collapse supernovae}
\label{sec:ccrate}

About 37\% of the cluster ($D<1.5$ Mpc) SNe (16 out of 44) are CC, 
and this allows a 
first measurement of the CC rate in galaxy clusters.
As shown in panels b 
of Figs.~\ref{fig:massrate} and \ref{fig:brate}, the cluster CC SNe 
are hosted by the cluster
spirals and irregular galaxies. We do not detect any
variation of the CC rate related to the environment, indicating
a similar IMF in cluster and field galaxies.

It should be noted that two effects 
are present that could hide any intrinsic differences: 
1) our determination of galaxy membership is not perfect, as discussed in
Sect.~\ref{sec:rate}, and 
2) in clusters, 
the number of CC SNe per galaxy type is small,
and only large differences, of the order of 50\% or more,
could be significantly detected.


\section{Type Ia supernovae}
\label{sec:iarate}

Type Ia SNe are present in all types of galaxies. As shown by \citet{M05},
their rate per unit mass sharply rises from ellipticals to
irregulars by a factor of $\sim$20, 
and from red galaxies to blue galaxies by a factor of $\sim$30. 
This is a strong indication
that a significant fraction of these explosions are due to young systems,
tracing the SFR.
Figure~\ref{fig:massrate} shows that this trend is present in both 
environments, with cluster and field galaxies showing similar behaviours. 

Early-type (E/S0) clusters galaxies ($D<0.5$ Mpc), comprising 3 times fewer 
early-type galaxies than the field, and 1.6 times less ``sensitivity''
(i.e., the product of stellar mass times control time),
contain 2.2 times more SN Ia events (11 vs. 5).
As a consequence, clusters early-types have a higher rate
($0.066^{+0.027}_{-0.020}$ SNuM) 
than field early-type galaxies ($0.019^{+0.013}_{-0.008}$ SNuM).
The difference is present both in the rates per unit mass 
(Fig.~\ref{fig:massrate})
and per unit B luminosity (Fig.~\ref{fig:brate}).

The statistical significance of the rate difference can be estimated in several
ways. One way is to apply the $\chi^2$ test by 
considering that the numbers of detected SNe are affected by Poisson errors. 
Excluding the galaxies at intermediate distances ($0.5<D<1.5$~Mpc), 
the null hypothesis, i.e., the hypothesis of no rate
difference between cluster and field, would predict 39\% of the 16 SNe 
(i.e., 6.2 SNe instead of 11)
in clusters and 61\% (9.8 instead of 5) in the field.
For one degree of freedom,
we obtain that the statistical significance of the difference is 97.7\%.

A second way to compute the statistical significance of the difference
in rates is to consider the binomial distribution of the probability
of a SN to explode in an early-type galaxy either in the cluster or in the
field. The probability of detecting 11 SNe or more 
in clusters out of 16 events in total, when the null-hypothesis of
expectation is 0.39, is  1.3\%. In other words,
we can exclude the null hypothesis that there is no difference
in rates between the two populations with a
confidence level higher than 98.7\%, in good agreement with the previous method.

A similar result is obtained when splitting the galaxy sample according
to the galaxy (B--K) color, as in \citet{M05},
rather than by morphological type.
Red galaxies with (B--K)$>3.7$, corresponding to ellipticals and red early-type 
spirals, have larger rates in clusters 
($0.047^{+0.016}_{-0.012}$ SNuM)
than in the field ($0.029^{+0.011}_{-0.008}$ SNuM). 
This difference has a lower statistical significance ($\sim 1.5
\sigma$), the reasons for which are explained in
Sect.~\ref{sec:discussion}.

\smallskip
The presence of a higher rate in cluster early-type galaxies is not strongly
dependent on the value of the cluster membership threshold
distance $D$. Figs~\ref{fig:massrate} and \ref{fig:brate} show that the
rate excesses corresponding to $D$=0.5~Mpc and 1.5~Mpc are very similar.
The statistical significance also remain very similar. 
For $D=1.5$ Mpc, cluster galaxies contain 15 out of 20 SNe and 49\% of the
sensitivity, resulting in a binomial significance of $>$98\%.
Even for very small values of $D$ the effect remain
present: using $D=0.2$ Mpc, the cluster rate (based on only 5 SNe) is 0.048
SNuM, to be compared with the field rate of 0.019 SNuM.

In conclusion, the difference is statistically significant but not at
a level above any doubt. Also, because of the possible presence of
low-level systematics,  a larger number of SNe 
in a more homogeneous sample of galaxies is needed to
confirm this effect.

\smallskip

We note that our cluster SN rate does not take into account the
possible contribution from intergalactic cluster SNe \citep{galyam03,maoz05}, 
which would be missed by 
the searches in the \citet{C99} sample as they were targeted at 
single galaxies rather than large fields.
\citet{galyam03}
quantified this possible contribution at $\sim$20\% of the total cluster rate
(see also \citealt{sand07}). 
This extra rate should be added to
the cluster rate but not to the field one, and would increase
the measured difference
and its statistical significance.

\begin{figure}  
\centering
\includegraphics[width=9cm]{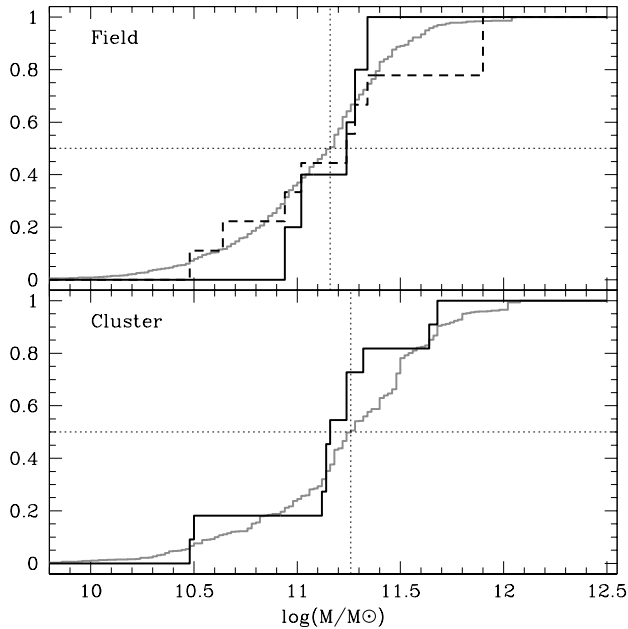}
\caption{
Cumulative distributions of SNe (in black) and sensitivity 
(i.e., the product of control time and stellar mass, in grey) 
as a function of the stellar mass for field (top panel) and cluster galaxies
(bottom panel). 
In the top panel, the solid black line shows the field SNe,
while the dashed one includes the SNe in galaxies at intermediate 
distances $D$, to increase the size of the sample.
The vertical dotted lines show the masses corresponding to
half of the distributions.
In both cases the SN distributions are compatible with the distributions of the
sensitivity.
\label{fig:masshist}
}
\end{figure}

\section{Dependence of the rates on mass, morphology and radio power}
\label{sec:mass}

Before discussing the origin of the type Ia
rate difference between clusters and field,
we explore the possibility
that the rates depend on some other galaxy parameter,
and that the dependence with environment is only an indirect effect.
We consider three possible parameters that are known to be important for the
evolution of galaxies: ({\em i}) stellar mass; ({\em ii}) morphology ;
({\em iii}) colors; and
({\em iv}) radio power of the parent galaxies.

\begin{figure}  
\centering
\includegraphics[width=8cm,height=6cm]{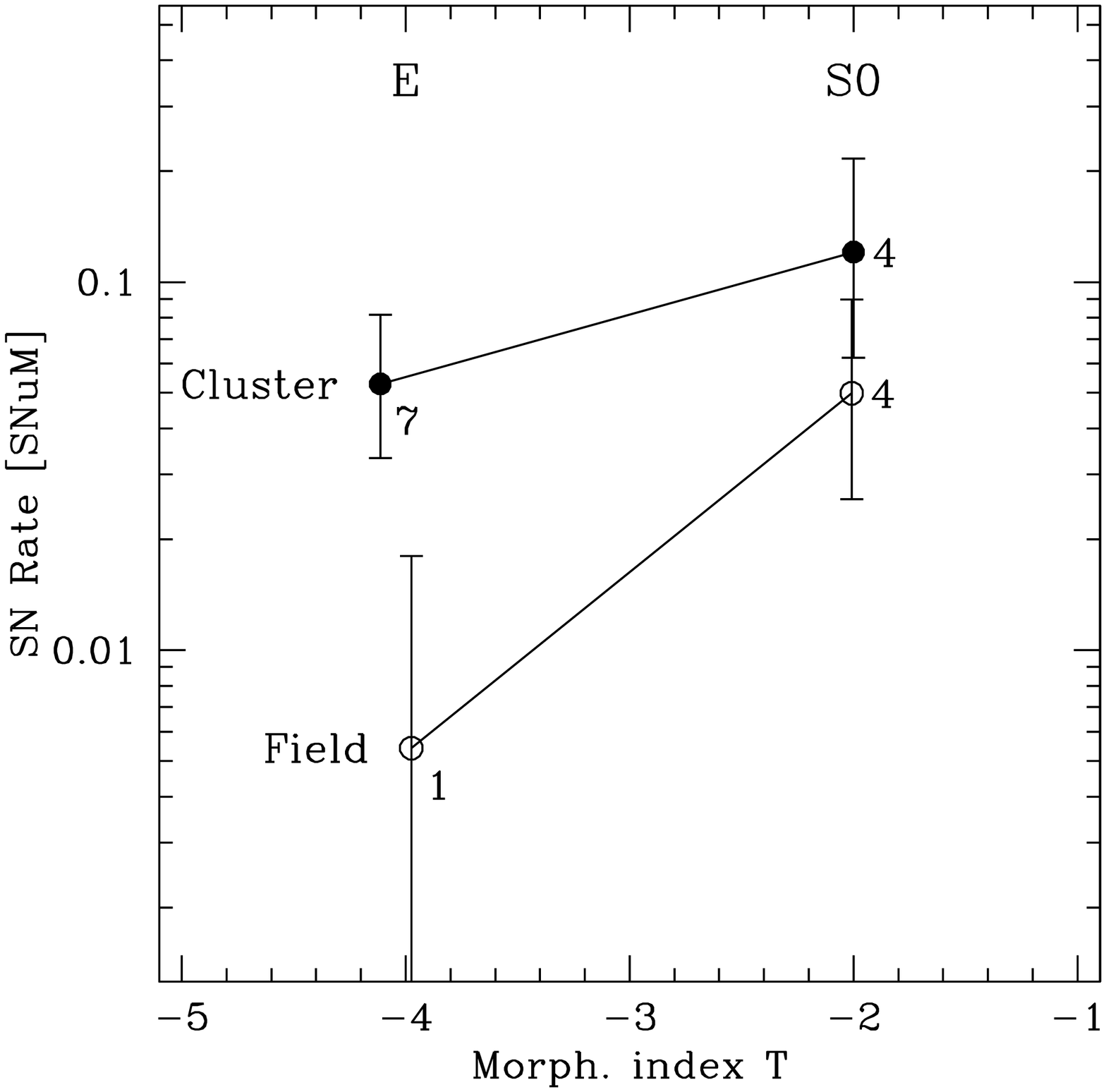}

\includegraphics[width=8cm,height=6cm]{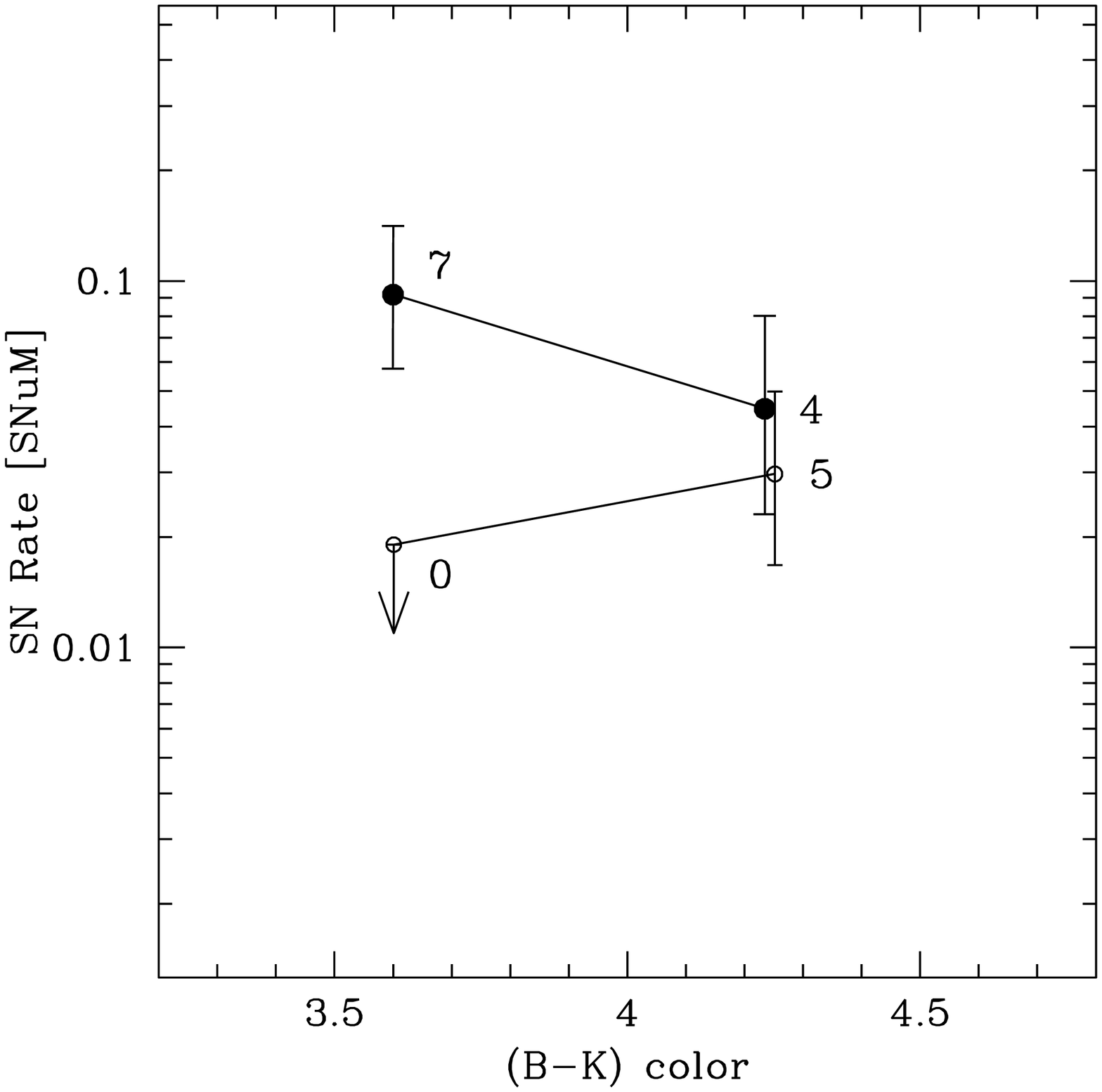}

\includegraphics[width=8cm,height=6cm]{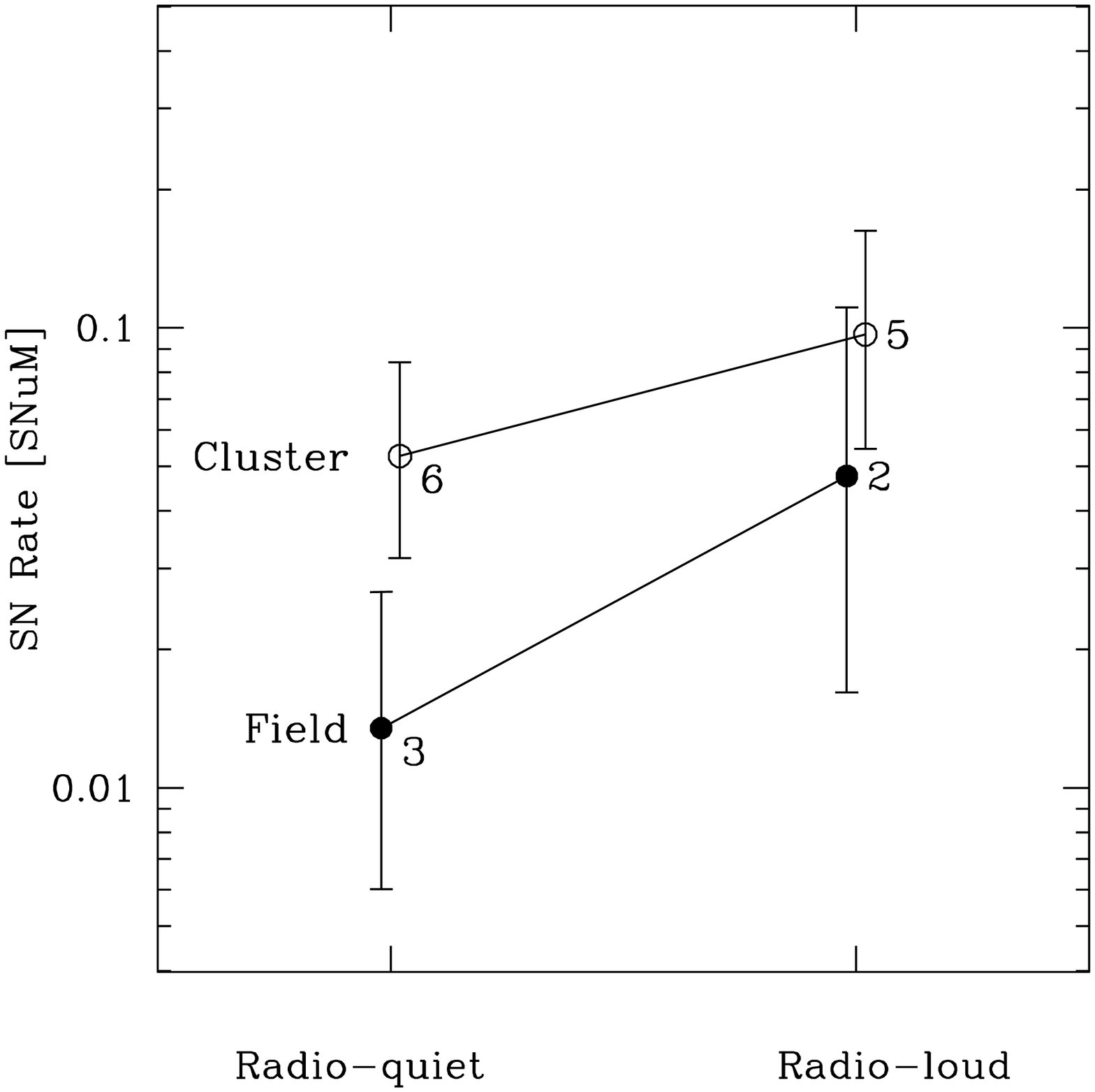}
\caption{
{\em Top panel}:
Type Ia SN rate in cluster and field early-type galaxies
as a function of
the morphological index T, separating ellipticals and S0s.
Next to each point, the number of corresponding SNe is reported.
Both types of galaxies have higher SN Ia rates in cluster than in the field,
and S0 galaxies show higher rates than ellipticals
both in cluster and in the field.
{\em Central panel}: 
as above, as a function of the (B--K) color of the parent galaxies.
The arrow shows a 1-sigma upper limit.
The difference in SN rate appear to be related to blue galaxies.
{\em Central panel}: 
as above, as a function
of the radio power of the parent galaxy, separating radio-quiet
and radio-loud galaxies according to the definition in \citet{dellavalle05}.
Radio-loud galaxies appear to have higher rates both in clusters and in the
field.
\label{fig:checks}
}
\end{figure}

\begin{enumerate}

\item Mass. Following early findings by \citet{gavazzi93} and \citet{cowie96}, 
``downsizing'' has recently become the standard paradigm to describe galaxy
formation (see \citealt{renzini06} for a review). 
According to this scheme, supported
by numerous observations, galaxies with different masses follow different
evolutionary paths. Therefore, it is interesting to study how the possible
rate difference described in the previous section depends on galaxy mass.
This test is also very useful to detect any bias related to the mass,
i.e., to the fact that it is more difficult to detect SNe in more massive
galaxies, with the higher stellar backgrounds that they pose for SN surveys.
In our case, mass does not appear to be at the origin of the difference.
As shown in Fig.~\ref{fig:masshist}, 
plotting the the cumulative distributions of sensitivity and SNe 
as a function of host galaxy mass,
cluster and field galaxies have 
similar mass distributions, with differences limited to below 10\%. 
Furthermore, the distribution of the number of detected SNe with galaxy mass 
follows that of the ``sensitivity'', i.e., galaxies of different mass
have the same SN rate per unit mass.
Applying the KS test of these two distributions, the resulting 
normalized ``D'' values
are 1.18 and 1.14 for clusters and field, respectively, i.e., in both
cases the distributions are fully consistent with being extracted from the
same sample.

\item Morphology. S0 galaxies are known to have younger stellar populations than
ellipticals (see, for example, \citealt{mannucci01}).  
As a consequence, different mixes of morphological
types in clusters and in the field could be at the origin of the 
different rates.
The top panel of figure~\ref{fig:checks} shows that this is not the case.
Both S0 and elliptical galaxies, independently, show similar higher rates
in clusters, with most of the observed difference due to ellipticals. 
Also, 
our sample has similar fractions of both types of galaxies
in clusters and in the field: ellipticals are 63\% of the cluster 
early-type galaxies  and 56\% of the field early-type galaxies.

\item Color. The (B--K) color of early-type galaxies 
is very sensitive to metallicity,
which could have an important effect on SN rate. 
Also, the presence of traces of star formation could produce bluer colors.
Adding a burst producing 0.1\% of new stars in $10^8$ yr 
can change the (B--K) color by up to 0.2 mag. 
The presence of dust, of multiple subsequent bursts, and differences 
between the timescales
of starburst evolution and SN explosion, can reduce the color difference
between galaxies with and without SNe to a much lower level but, possibly,
not completely remove it.
It is therefore interesting to study if the enhancement of the SN rate 
is related to the (B--K) color of the parent galaxies.
In our sample, the B band photometry is from the RC3 catalog 
\citep{devac91} and
near-IR K band photometry from 2MASS \citep{jarrett03}.

The central panel of figure~\ref{fig:checks} 
suggests that most of the difference in SN rate is due to blue galaxies, 
i. e., those having (B--K)$<$3.9. These galaxies 
contain 7 SNe in clusters and none in the
field, even though in our sample we have similar sensitivities 
in the two environments.
This would be consistent with the interpretation of the SN excess as due to
a higher level of recent star formation in cluster early-type galaxies.
However, such a result is {\em not} confirmed by Sloan Digital Sky Survey 
(SDSS) photometry in the $g$ band 
(centered at 4686\AA, similar to the center of the Johnson B band at
4400\AA) and in the $u$ band (centered at 3551\AA), available for
about 1/4 of the total sample. No difference is seen in the
distributions of ($g$--K) and ($u$--K) colors between cluster and 
field galaxies. 
Also, a systematic change
in the (B--$g$) color is measured as a function of the apparent B magnitude,
with variations of the order of 0.1 mag. This trend depends on both 
environment and luminosity,
with larger differences found in cluster galaxies and in faint objects.
The reason for this is not clear, but is probably due to contamination
of the RC3 magnitudes by the light of nearby galaxies.
Unfortunately, many galaxies
hosting SNe are partly saturated in the SDSS images, 
and their colors cannot be accurately measured.

As a consequence, the dependence of the SN rate on galaxy colors in
Fig.~\ref{fig:checks} should be
taken with caution, because our photometry is not precise enough to
unambiguously detect variations at the 0.1 mag level.

\item Radio properties. \citet{dellavalle05} demonstrated that radio-loud 
early-type galaxies have higher rates of SNe Ia
than radio-quiet galaxies of the same type.
The overproduction of SNe is explained as the result of
a residual activity of star formation, produced by recent episodes 
of merging or gas accretion (see also \citealt{dellavalle03}). 
The alternative mechanism proposed by \citet{livio02} to explain
the overproduction of novae in M87, i.e., 
Bondi accretion of the material of the radio jet,
cannot be invoked here because SNe require mass accretions larger by 4 order
of magnitudes \citep{dellavalle05}.
In principle, the higher rates in clusters could be due to 
a higher fraction
of radio-loud galaxies in clusters than in the field. To check this possibility
we have studied the rates in the two environments after splitting 
the sample into radio-loud and radio-quiet galaxies.
The bottom panel of Fig.~\ref{fig:checks} shows that the rate 
difference cannot be related to the radio power because both
radio-loud and radio-quiet galaxies have higher SN rates in clusters
than in the field. Also, the fraction of radio-loud galaxies in both samples
is similar, being 16\% in clusters and 12\% in the field.
It appears that both properties (radio-power and
environment) are, separately, affecting the rates. This is even more
evident when comparing radio-quiet field galaxies with radio-loud cluster
galaxies, the latter having rates 20 times larger than the former.

\end{enumerate}

In conclusion, the dependence of the SN Ia rate on the environment is not
due to the other parameters considered here, and seems an independent
effect.

\smallskip

Also, we analyze several possible selection effects or observational biases
that could be at the origin of the difference:

\begin{enumerate}
\item The host galaxies are the same in the two samples, i.e., 
they were selected
on the basis of only the morphological index T. As a consequence,
no strong differential detection efficiencies are expected to be present
in cluster and field samples

\item Dust is not expected to be a major problem in these early-type galaxies
and, in particular, it is not supposed to give an important differential
effect. \citet{maiolino02}, \citet{mannucci03,mannucci07} 
and \citet{cresci07} have shown that
dust corrections to the rates can be important but only for the
very dusty starburst galaxies dominating the star formation density at high
redshifts.

\item some remaining observational biases could be related to the fact 
that the the \citet{C99} sample makes use of 5 visual 
and photographic searches (see \citealt{C97} for details). 
A spurious effect could be produced if different searches target significantly
different fractions of field and cluster galaxies, and if 
inhomogeneous estimates of the sensitivities are present.
Even though we cannot check for 
the presence of these two problems, we do not think they produce
a dominant effect, mainly because the fraction of cluster and field galaxies 
is expected to be similar in all the surveys.
\end{enumerate}

In conclusion, we do not identify any selection effect 
that can be responsible for the observed difference.
If the difference is not a pure
statistical fluctuation, it must related to
environmental differences.

%
\section{Discussion and conclusions}
\label{sec:discussion}

The interpretation of the possible difference in SN Ia rate
between cluster and field early-type galaxies is not straightforward.
As the observed rate is the convolution of the SFH with the DTD, the
differences could be due to either of these functions.

\begin{enumerate}

\item The first possibility is that the rate difference is due to 
differences in the stellar populations.
\citet{M05}, \citet{sullivan06}, and \citet{aubourg07}
have shown that the type Ia SN rate has a strong dependence on the 
parent stellar
population, with younger stars producing more SNe. 
The difference in SN rate could be related to this effect, i.e., 
to a higher level of recent star formation in cluster ellipticals.
Only a very small amount of younger stars is needed, because
the amplitude of the DTD 
at short times can be hundreds of times larger than at
long times.
As an example, the \citet{greggio83} single-degenerate model
has 300 times more amplitude at $10^8$ yr than at $10^{10}$ yr, and this means
that a recently formed stellar population contributing 0.3\% of the mass
can provide
as many SNe as the remaining 99.7\% of old stars. For the ``two channel''
model by \citet{mannucci06}, the amount of young stars needed can be even lower,
at the 0.1\% level, as this DTD amplitude ratio between $10^7$ and $10^{10}$ 
years is as large as 1000.  

\smallskip

The presence of traces of star formation in early-type galaxies
is not inconsistent with other observations.
Many ellipticals show signs of recent interactions
or star formation activity: faint emission lines \citep{sarzi06},
tidal tails \citep{vandokkum05}, dust lanes \citep{vandokkum95,colbert01},
HI gas \citep{morganti06}, molecular gas \citep{welch03}, and
very blue UV colors \citep{kaviraj06,schawinski07,haines07,kaviraj07}. 
Even if the interpretation of most of these effects is matter of debate
(for example, \citealt{serego07} have found only small amounts of HI gas 
in cluster ellipticals), the observations suggest a widespread, 
low-level presence of star formation.

The dependence of this presence with environment is not settled yet.
\citet{ferreras06} have found
evidence for recent star formation, at the percent level, in ellipticals
in compact groups, but not in field ellipticals. In contrast,
\citet{verdugo07} and \citet{haines07}
have found higher levels of 
present star formation in field rather then cluster early-type galaxies.

\smallskip

Some studies (see, for example, 
\citealt{sanchez06}, \citealt{bernardi06} 
and \citealt{collobert06}), 
have found younger ages in field early-type galaxies 
with respect to cluster galaxies 
(but \citealt{serego06} have found no difference).
Taken at face value, this would seem
to contradict the star formation interpretation of the SN rate, 
but this is not necessary the case.
Field ellipticals could be younger that cluster ellipticals,
but nevertheless they could show a lower level of {\em present} star
formation. 
The difference in the 
age of the {\em dominant} stellar population of early-type galaxies,
of the order of 1 Gyr for ages of about 12 Gyr,
might not be directly related to
the amount of star formation in the last few $10^8$ years.
Such a contribution cannot be detected in the integrated colors
of the galaxies. The expected differences are at the 0.05 mag level
for the (B--K) color, assuming the younger stars are not associated with dust, 
and even smaller (0.02 mag for $A_V$=1), allowing for dust extinction.

It is usually assumed that early-type galaxies can form new stars only after
merging with a small, gas rich galaxy, because usually they do not 
host much interstellar gas. The average amount of stars formed
is proportional to the merger (or encounter) rate, to the typical amount of 
gas in the accreted galaxy, and to the efficiency of star formation 
in the accreted gas.
It is possible that one or more of these
quantities are larger for cluster galaxies than for field galaxies because of
the different galaxy volume density and galaxy-galaxy encounter velocity.

\smallskip

If this is the correct interpretation, the ``prompt'' population of 
SNe Ia would be associated with the explosion of CC SNe from the same young
stellar populations.
If a SN Ia is to explode within $10^8$ yr of the formation of its progenitor, 
the primary star of the progenitor binary system 
must have a mass above 5.5 \msun\ to allow for the 
formation of a white dwarf in such a short time. 
\citet{mannucci06} have shown that reproducing 
the observed SN rates by using the ``bimodal'' DTD in that paper
implies that about 7\% of all stars between 5.5 and 8 \msun\ 
explode as ``prompt'' SNe Ia, while the ``tardy'' population corresponds to 
a lower explosion efficiency, about 2\%, and on a much longer timescale
(see also \citealt{maoz07} for various estimates of these efficiencies).
For a Salpeter IMF and assuming that 100\% of the stars between 8 and 40 \msun\ 
end up at CC SNe, we expect 1.3 CC SNe for each ``prompt'' type Ia.
Assuming that the difference between cluster and field early-type galaxies is
due to the ``prompt'' SNe Ia, the rate of this population is of the order of 
0.066-0.019=0.047 SNuM (see Table~\ref{tab:massrate}). 
Converting this rate to an observed number, 
about 2 CC SNe are expected in the cluster early-type galaxies
of our sample, consistent with our null detection at about 1.3$\sigma$
level.
We conclude that the non detection of CC in the early-type galaxies
belonging to our sample
and the corresponding upper limits to the CC rate are
consistent with the hypothesis of a ``prompt'' Ia component.
We also note that some CC SNe have been discovered in the
recent past in prototypical early-type galaxies.
\citep{pastorello07}.

\item A second possible interpretation is that the higher rate in cluster 
early-type galaxies is related  
to differences in the DTD. If the stars in ellipticals 
are 9-12 Gyrs old (see, for example, \citealt{mannucci01}), the SN rate
is dominated by the tail of the DTD at long times. 
Differences in the environments could produce small differences in the
shape of this function, for example because of the higher numbers of
encounters.

A interesting possibility is also that the changes in the DTD are 
related to differences in metallicity
between cluster and field early-type galaxies, as discussed by
\cite{sanchez06,bernardi06,collobert06} and \cite{prieto07}. 
The differences between cluster and field galaxy metallicity 
presented by these
papers are neither large nor always in the same direction. Nevertheless
systematic, although not large, differences in metallicity could be present 
and produce significant changes in the
DTD, for example, by affecting the efficiency of mass loss during
the complex life of a binary system.

\end{enumerate}

Table~\ref{tab:history} lists the different measurements of the SN rate
in early-type cluster galaxies. The evolution of this rate can be
compared with the history of star formation of the parent galaxies to derive the DTD.
Currently published cluster SN rates at z$>$0.2 are
too uncertain to permit any strong conclusions.
However, 
current and future searches for SNe are expected to 
change this situation and allow for the derivation of meaningful 
constraints (see, for example, \citealt{sharon06}).

\smallskip

To summarise, we have used a sample of 136 SNe in the local universe
to measure the SN rate as a function of environment. For the first
time, we measure the CC SN rate in clusters. 
We find it is very similar to the CC SN rate in field galaxies, 
suggesting that the IMF is not a strong function of the 
environment. For Ia SNe, the rates in clusters and in the field
are similar for all galaxy types except for the early-type systems, 
where we detect a significant excess in clusters. 
This excess is not related to other properties
of those galaxies, such as mass, morphology, or radio loudness.
Environments itself 
appears to be important. We interpret this effect as possibly
due to galaxy-galaxy
interaction in clusters, either producing a small 
amount of young stars (of the order of the percent in mass over one 
Hubble time),
or affecting the evolution of the properties of the binary systems.

\bigskip

{\bf Acknowledgments}

We thank Sperello di Serego and the MEGA group (Arcetri Extragalactic Meeting)
for useful discussions about the properties of
elliptical galaxies.
This research has made use of the NASA/IPAC Extragalactic Database (NED) 
which is operated by the Jet Propulsion Laboratory, 
California Institute of Technology, 
under contract with the National Aeronautics and Space Administration.
DM, MD, and AG thank the Kavli Institute for Theoretical Physics
for its hospitality.
This research was supported in part by the National Science
Foundation under Grant No. PHY05-51164.


\end{document}